\documentclass[a4paper,11pt]{article}
\pdfoutput=1 

\usepackage{jinstpub} 
\usepackage{graphicx}
\usepackage{subcaption}

\title{\boldmath Status of the laboratory infrastructure for detector calibration and characterization at the European XFEL}

\author[a,1]{N. Raab,\note{Corresponding author.}}
\author[a]{K.-E. Ballak,}
\author[a]{T. Dietze,}
\author[a]{M. Ekmedzi\u{c},}
\author[a]{S. Hauf,}
\author[a]{F. Januschek,}
\author[a]{A. Kaukher,}
\author[a]{M. Kuster,}
\author[a]{P. M. Lang,}
\author[a]{A. M\"{u}nnich,}
\author[a]{R. Schmitt,}
\author[a]{J. Sztuk-Dambietz}
\author[a]{and M. Turcato}

\affiliation[a]{European XFEL GmbH,\\Holzkoppel 4, Schenefeld, Germany}

\emailAdd{natascha.raab@xfel.eu}

\abstract{
The European X-ray Free Electron Laser (XFEL.EU) will provide
unprecedented peak brilliance and ultra-short and spatially coherent
X-ray pulses in an energy range of $0.25$ to $25$ keV. The pulse
timing structure is unique with a burst of 2700 pulses of $100$ fs
length at a temporal distance of $220$ ns followed by a $99.4$ ms gap.
To make optimal use of this timing structure and energy range a great variety of
detectors are being developed for use at XFEL.EU, including 2D X-ray
imaging cameras that are able to detect images at a rate of 4.5 MHz,
provide dynamic ranges up to $10^5$ photons per pulse per pixel under
different operating conditions and covering a large range of angular
resolution~\cite{requirements,Markus}. In order to characterize,
commission and calibrate this variety of detectors and for testing of
detector prototypes the XFEL.EU detector group is building up an
X-ray test laboratory that allows testing of detectors with X-ray
photons under conditions that are as similar to the future beam
line conditions at the XFEL.EU as is possible with laboratory
sources~\cite{infra}. A total of four test environments provide the
infrastructure for detector tests and calibration: two portable setups
that utilize low power X-ray sources and radioactive isotopes, a test
environment where a commercial high power X-ray generator is in use, 
and a pulsed X-ray/electron source which will provide pulses as short
as 25 ns in XFEL.EU burst mode combined with target anodes of
different materials.  The status of the test environments, three of
which are already in use while one is in commissioning phase, will be
presented as well as first results from performance tests and
characterization of the sources.}


\arxivnumber{} 

\proceeding{18$^{\text{th}}$ International Workshop on Radiation Imaging Detectors\\
  3-7th July 2016\\
  Barcelona, Spain}

\begin{document}
\maketitle
\flushbottom
\section{Introduction}
\label{sec:intro}
\begin{figure}[htbp]
\centering 
\qquad
\begin{subfigure}{2.1in}
\includegraphics[width=1.\textwidth]{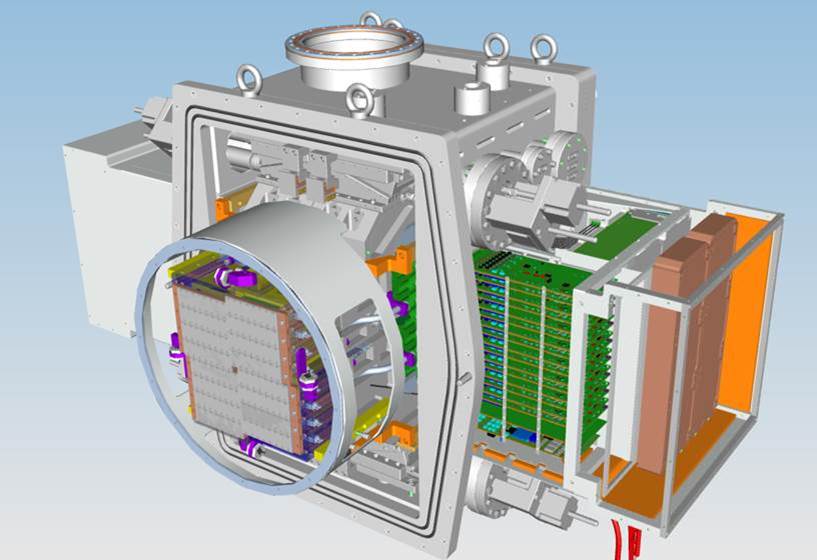}
\end{subfigure}
\hspace{2mm}
\begin{subfigure}{1.5in}
\includegraphics[width=1.\textwidth]{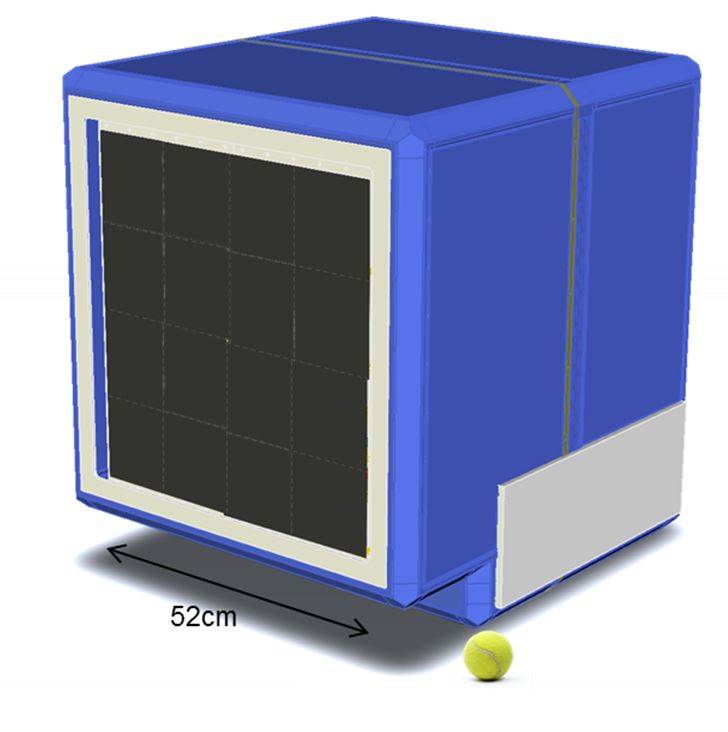}
\end{subfigure}
\hspace{2mm}
\begin{subfigure}{1.7in}
\includegraphics[width=1.\textwidth]{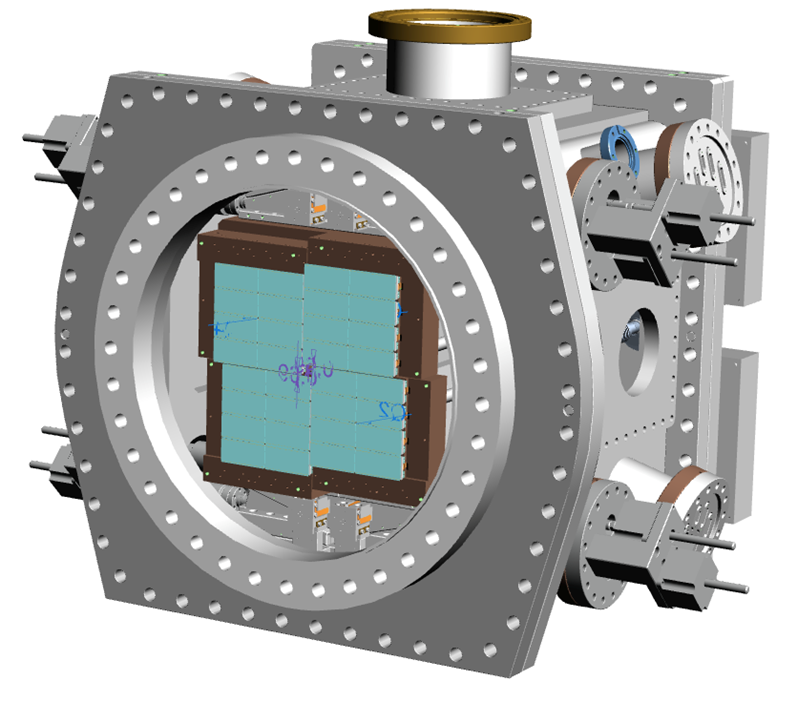}
\end{subfigure}
\caption{\label{fig:2Ddetectors}  From left to right 3D CAD models of the AGIPD (image is courtesy by the AGIPD consortium), LPD (image is courtesy by the LPD consortium) and DSSC (image is courtesy by the DSSC consortium) 2D imaging detectors are shown including their mechanical housing (LPD) or vacuum setup (AGIPD, DSSC).}
\end{figure}

The European X-ray Free Electron Laser is a high-brilliance fifth
generation X-ray light source located in the area of Hamburg. It will provide spatially coherent
X-rays in the energy range between $0$ and $25$ keV, delivered in 10 bursts of X-ray pulses per second, where each
burst consists of up to $2700$ pulses of $100$ fs length at a rate of
$4.5$ MHz~\cite{altarelli,beamprops}.  
Three 4.5 MHz high rate 2D X-ray imaging cameras (Adaptive Gain
Integrating Pixel Detector (AGIPD)~\cite{agipd1},  Large Pixel
Detector (LPD)~\cite{LPDHart,LPDKoch} and DEPFET Sensor with Signal
Compression (DSSC)~\cite{dssc1,dssc2,dssc3} ), see figure \ref{fig:2Ddetectors}) are being developed with the goal of acquiring single X-ray pulse images of 1~mega pixel size at the XFEL.EU. In
addition small area 2D imaging cameras like the
FastCCD~\cite{fastccd} and pnCCD~\cite{pnccd} will be used for imaging
applications at the $10$ Hz burst rate.  

In order to avoid beam time being used for characterization, commissioning and calibration a test laboratory has been established. The laboratory aims at testing detectors in conditions which are as close as possible to those in their final installations. The test setups developed must handle the large variety of detector differences (energy range, vacuum or air operation, pixel size, sensor area, acquisition rate, etc.) anticipated as defined by the following requirements.

Test setups are necessary that allow measurements of the detectors under ambient conditions as well as in vacuum at a pressure of $10^{-5}$ mbar. 
X-rays have to be provided at different energies between 0 and 25 keV and with different illumination properties: point like illumination of single pixels is necessary as well as flat field illumination of large sensor areas to test for crosstalk and to apply flat field corrections.  
To test properties of the high rate cameras that depend on timing like the optimization of integration time or pulse to pulse afterglow, an X-ray source is required that can provide pulses in a time structure similar to that of the XFEL. 
To meet these requirements, four complementary test setups are being developed.

Two small size portable setups have been developed. PHEOBE operates under high vacuum and attaches to the vacuum vessel of the detector.
Little Amber is built for tests of small detectors, like the LPD two-tile system, under ambient conditions. 
The setups are used for low intensity flat field and energy response measurements, whereby PHEOBE provides X-rays at lower and medium energies and Little Amber provides photon energies in the medium and higher range.  

The Big Amber setup is large and allows test of the 1~mega pixel 2D imaging cameras. This setup is used for higher intensity flat field illumination of large sensor areas and beam spot illumination down to 50 $\mu$m at higher photon energies. 

Finally, the PulXar setup whose X-ray source provides pulses of $50$ ns or shorter delivered at a rate of 4.5 MHz in a burst of 2700 pulses in $600 \mu$s, is needed for measurement of all high rate detectors and the setup has to be vacuum compatible and provide a wide range of photon energies. 

Table \ref{tab:i} summarized the main parameters of the four setups, which are described in more detail in the following chapters.
\begin{table}[htbp]
\centering
\caption{\label{tab:i} Parameters of the four different setups.  As energies the positions of the main characteristic lines that are observed in the spectra are given}
\smallskip
\begin{tabular}{|l|p{2cm}|p{3cm}|p{2.6cm}|p{1.3cm}|p{1.3cm}|}
\hline
Setup&Operating conditions&Energies (keV)& Intensity
(counts~s$^{-1}$sr$^{-1})$ &Pulse length &Spot size \\
\hline
\hline
Little Amber& Ambient & 5.9, 9.7, 20.2 &  $10^{8}$ for 20.2 keV & n.a.  &  n.a. \\
\hline
Big Amber & Ambient &  8, 17.5 & $10^{11}$ for 17.5 keV & 168 ns  &   50$\mu$m \\
\hline
PHEOBE &$> 10^{-6}$ mbar & 5.9 , 9.7  & $10^{7}$ for 9.7 keV & n.a.  &   n.a. \\
\hline
PulXar & $> 10^{-8}$ mbar & 0.3 - 17.5 & to be measured & 50 ns  &   n.a. \\
\hline
\end{tabular}
\end{table}\\
 
\section{Portable vacuum compatible test setup PHEOBE}
The vacuum compatible setup PHEOBE is operated at pressures down to 
$5\times10^{-6}$ mbar and is attached to the vacuum vessel of a
detector via a gate valve. 
A Programmable Logic Controller is used to control the vacuum pumps, pressure gauges,  valves and manipulators of the setup. 
\begin{figure}[t]
\centering
\begin{subfigure}{2.0in}
\centering 
\includegraphics[width=1\textwidth,]{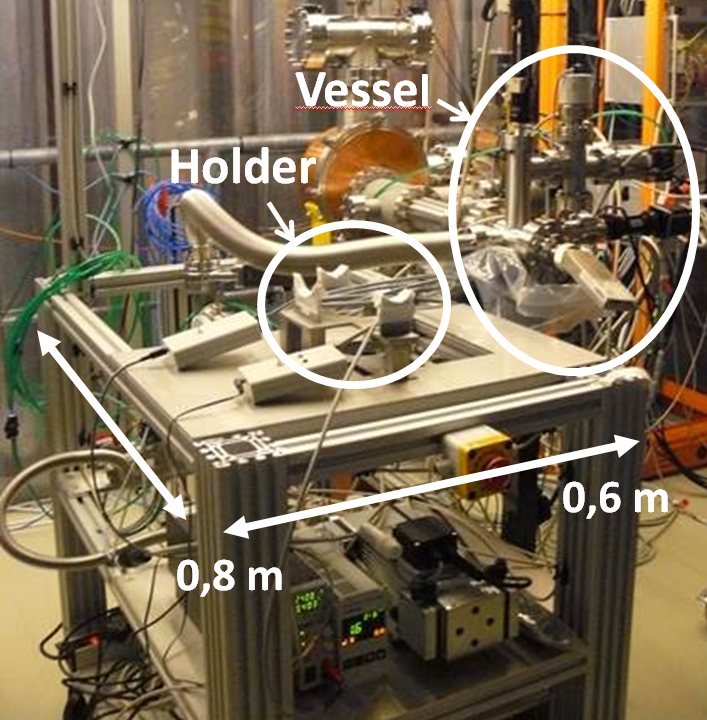}
\end{subfigure}
\hspace{3mm}
\begin{subfigure}{2.95in}
\centering 
\includegraphics[width=1\textwidth,]{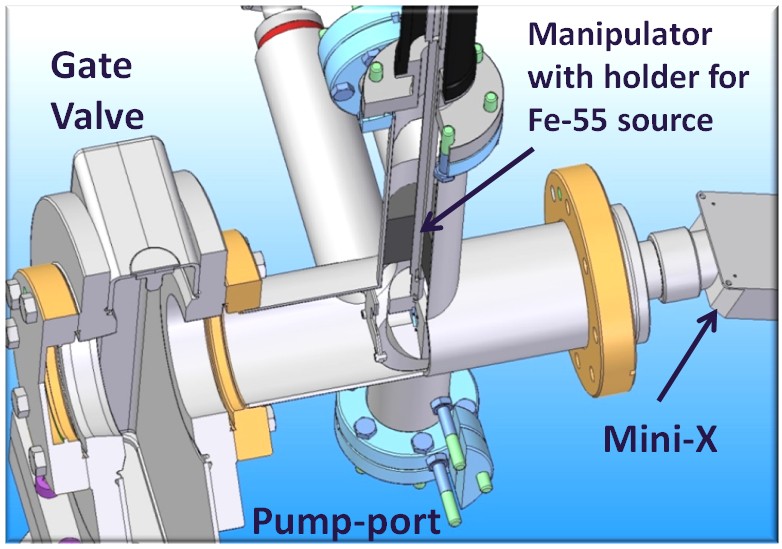}
\end{subfigure}
\caption{\label{fig:phe}Left: PHEOBE setup is shown, consisting of its pump system and vacuum vessel attached to the pnCCD detector vessel in the background. Right: CAD rendering of the vessel including the manipulator with source holder and Mini-X is shown. Copyright The Author(s), published under the CC-BY 4.0 license (\href{http://creativecommons.org/licenses/by/4.0/}{http://creativecommons.org/licenses/by/4.0/})}
\end{figure}
Figure \ref{fig:phe} shows the vacuum vessel of the setup, in which an Fe-55
source can be installed and moved by a linear translation stage in and out of the field of
view of the detector without breaking the vacuum. At the end of a
perpendicular manipulator a holder with filters attenuating the beam
can be attached. 
Additionally a low
power X-ray generator, an Amptek Mini-X \cite{minix} with Gold target, can be attached via a vacuum
flange.  The vacuum vessel provides shielding from X-rays up to $15$ keV. 

The emission spectra of the Fe-55 source, with an activity of $1.85$ GBq,  and the Mini-X are shown in figure~\ref{fig:las}.
As with all spectra in this publication, it was obtained with a
FastSDD detector from Amptek~\cite{fastsdd} (for the Fe-55 spectrum a highly transmissive low-energy window
\cite{windows} was used.)  
PHEOBE is a very flexible setup for providing X-rays at lower and medium energies under vacuum conditions.

\section{Portable ambient test setup Little Amber}
\label{littleA}

\begin{figure}[htbp]
\centering 
\qquad
\begin{subfigure}{2.1in}
\includegraphics[width=1.\textwidth,]{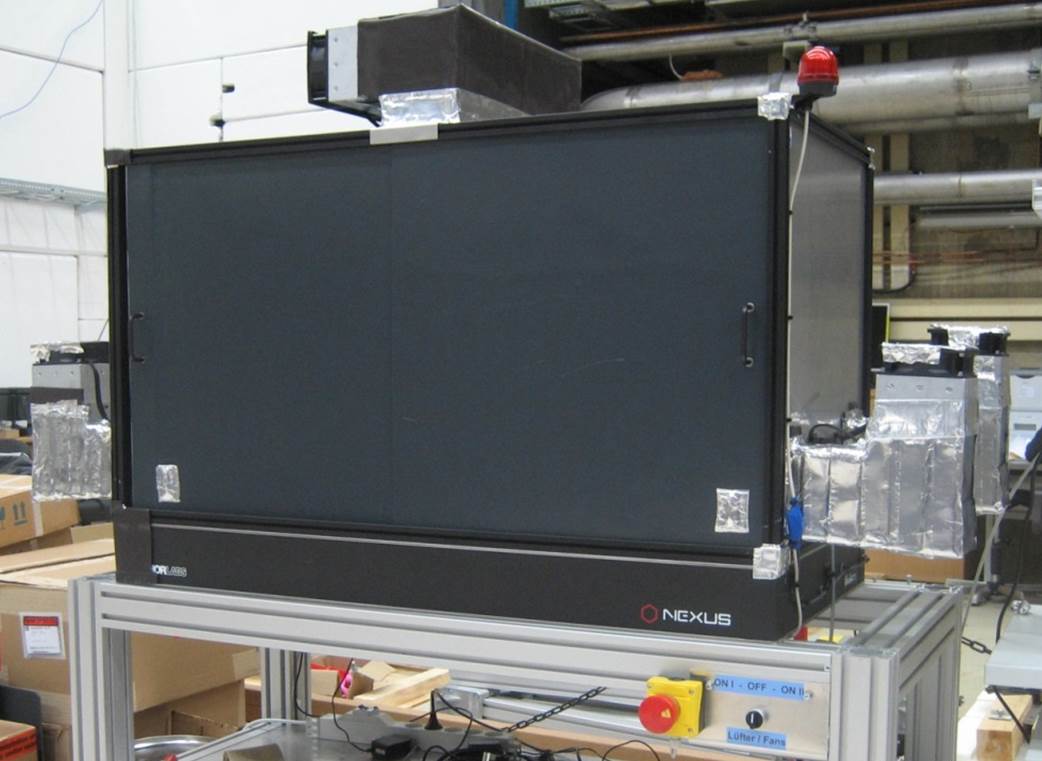}
\end{subfigure}
\hspace{2mm}
\begin{subfigure}{3.4in}
\includegraphics[width=1.\textwidth,]{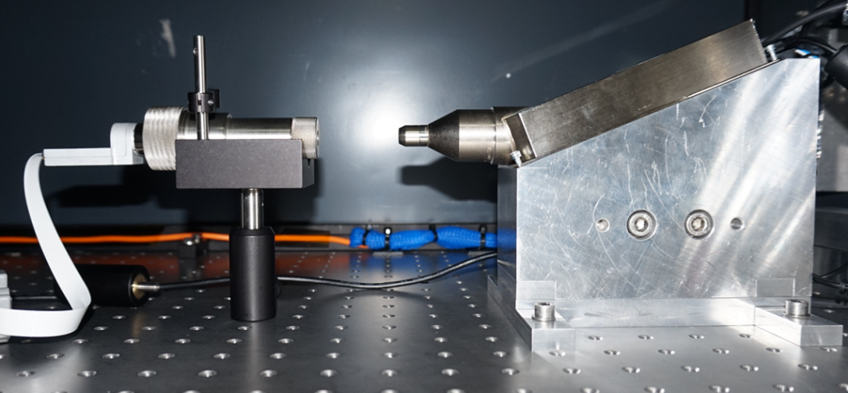}
\end{subfigure}
\caption{\label{fig:la} Left: the shielding cabinet of Little Amber with chicanes for ventilation. Right: a view inside showing the FastSDD detector mounted in front of the low power X-ray generator (Mini-X). Copyright The Author(s), published under the CC-BY 4.0 license (\href{http://creativecommons.org/licenses/by/4.0/}{http://creativecommons.org/licenses/by/4.0/})}
\end{figure}

\begin{figure}[htbp]
\centering
\includegraphics[width=0.49\textwidth,]{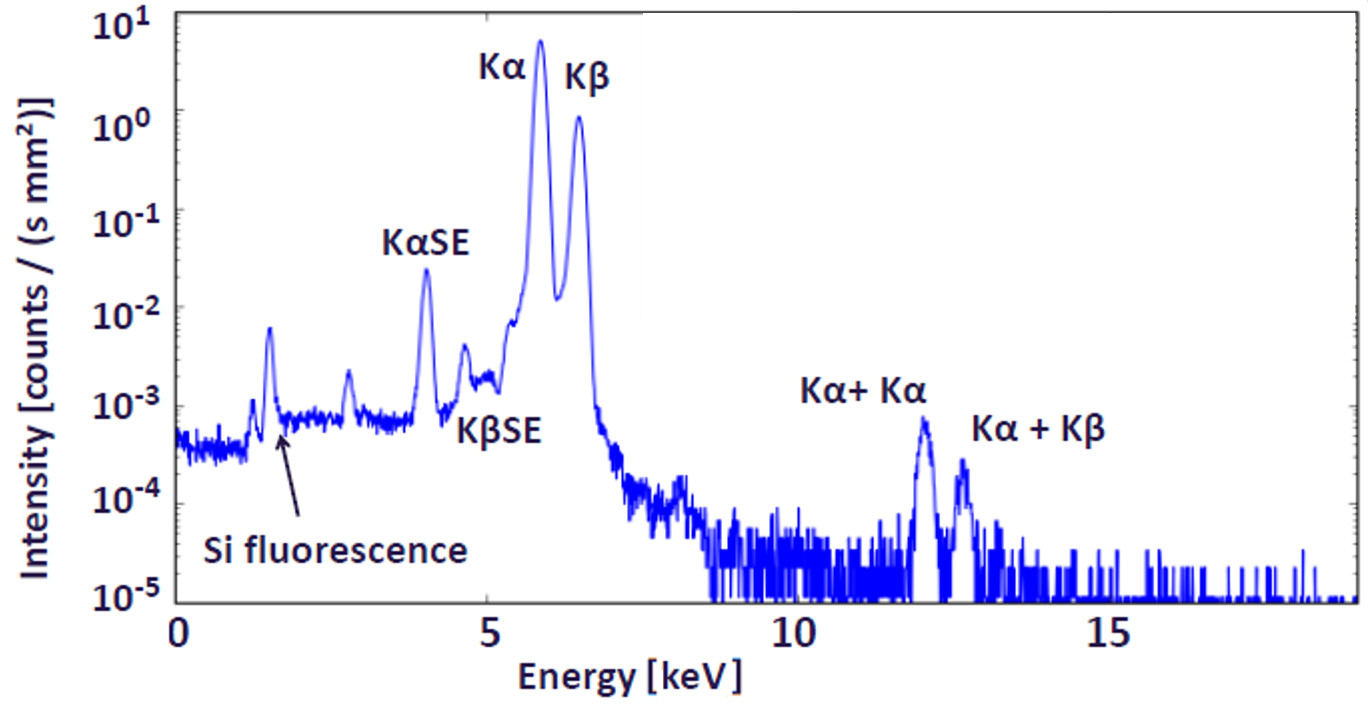}
\includegraphics[width=0.49\textwidth,]{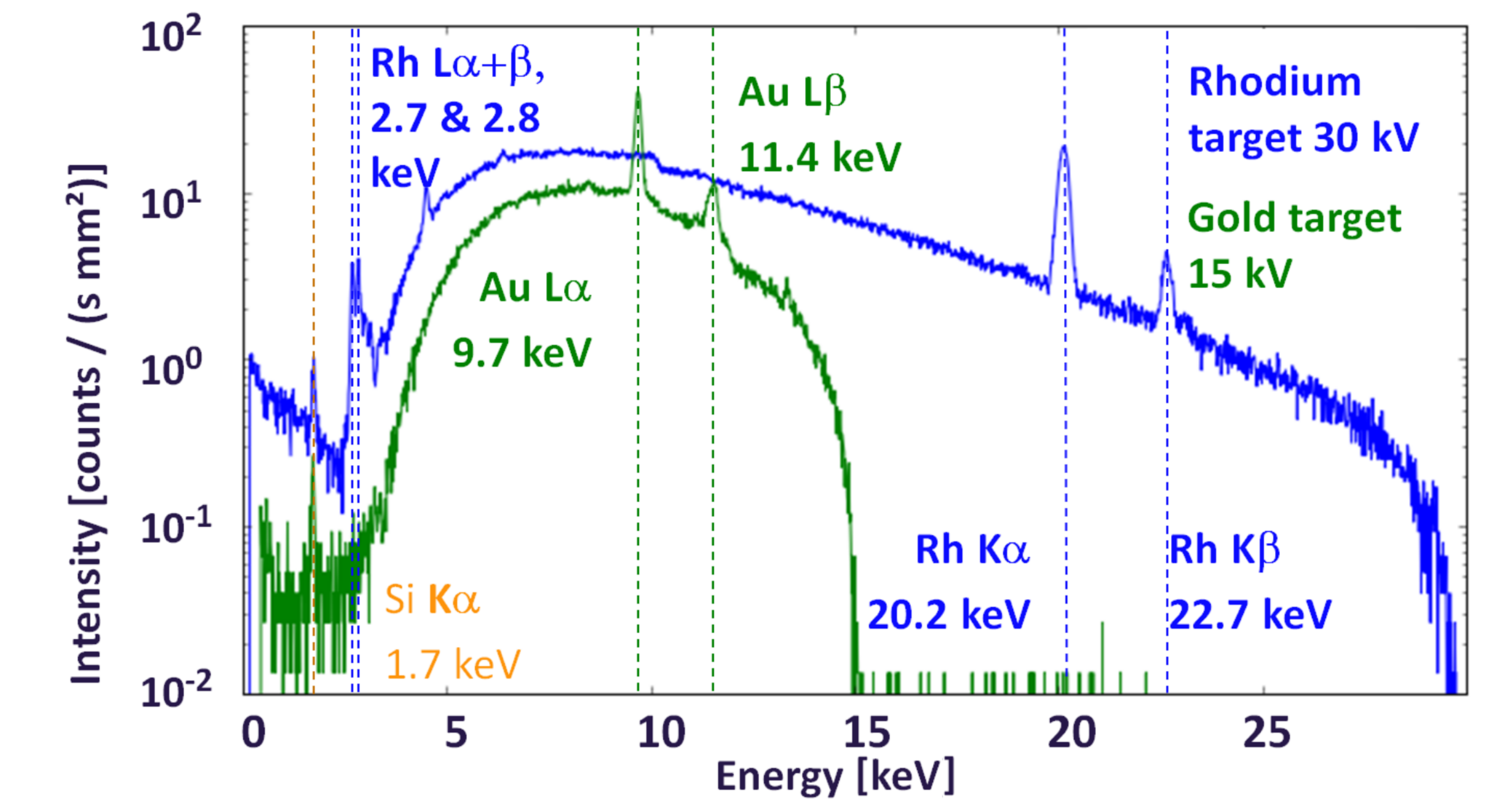}
\caption{\label{fig:las}Left:  X-ray emission spectrum of the Fe-55 source.  As indicated the characteristic Mn K$_\alpha$ and $K_\beta$ lines at $5.9$ and $6.5$ keV can be observed, as well as the pile up peaks at the energies $E(K_\alpha) + E(K_\alpha)$ and $E(K_\alpha) + E(K_\beta)$), the Si escape peaks at (K$_\alpha$SE and K$_\beta$SE )and the fluorescence peak of the silicon detector material. Right:  X-ray emission spectrum of the two Mini-X, that can be used in the Little Amber setup. For the Gold target characteristic X-ray L emission lines are visible, for the Rhodium target K- as well as the L-emission lines of the anode material can be observed. Both spectra show the K$_{\alpha}$ peak of the silicon
material of the detector. Copyright The Author(s), published under the CC-BY 4.0 license (\href{http://creativecommons.org/licenses/by/4.0/}{http://creativecommons.org/licenses/by/4.0/})}
\end{figure}

The Little Amber setup is shown in figure~\ref{fig:la}, its shielding cabinet is 60 cm $\times$ 120 cm $\times$ 60 cm large, such
that it can house detectors with sizes of up to approximately $0.5 $ m $\times$ 0.5 m $\times$ 0.5 m.
Inside the cabinet detectors are mounted on a breadboard in
front of the source and forced ventilation is possible for prototypes with
strong heat dissipation. The lead shielding of $1.5$ mm thickness
allows operation of the Amptek Mini-X tubes with accelerating voltages
up to $50$ kV. In addition to the X-ray tube with Gold target emitting
characteristic X-rays of the L$_{\alpha}$ and L$_{\beta}$ transitions
at $9.7$ keV and $11.4$ keV, a Mini-X with Rhodium target (see figure~\ref{fig:las}) can be used
which emits characteristic K$_{\alpha}$ and K$_{\beta}$ radiation at
$20.2$ keV and $22.7$ keV. For the measurement a FastSDD detector with a low-energy but light tight
window was mounted at a distance of $18$ cm in front of the Mini-X as
can be seen in figure~\ref{fig:la}. 

Little Amber is extensively used for basic tests of small detectors or small prototypes of the 2D cameras. 

\section{Ambient setup for 1~mega pixel detectors Big Amber}
\begin{figure}[htbp]
\centering 
\qquad
\begin{subfigure}{3.0in}
\includegraphics[width=1.\textwidth,]{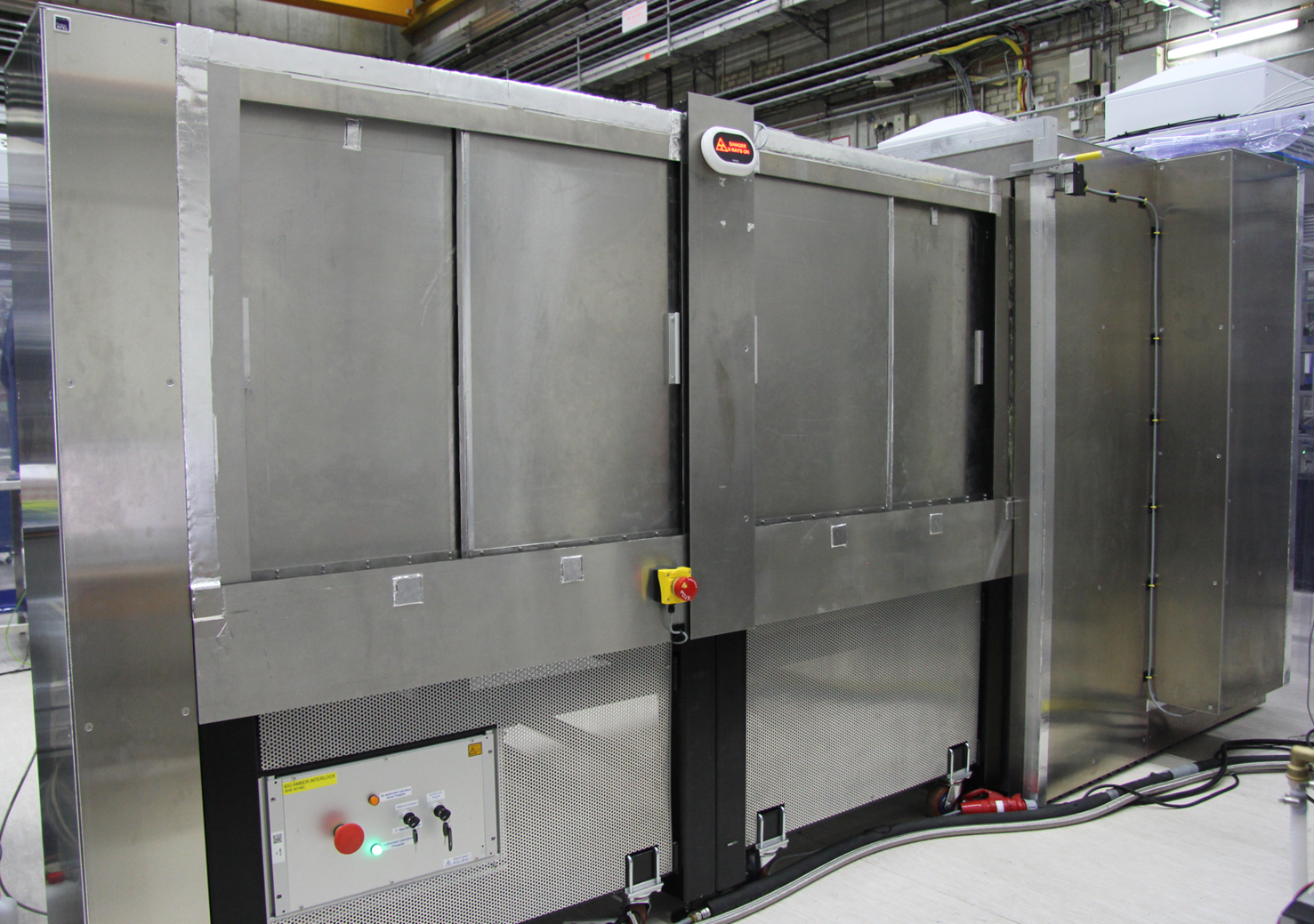}
\end{subfigure}
\hspace{5mm}
\begin{subfigure}{2.0in}
\includegraphics[width=1.\textwidth,]{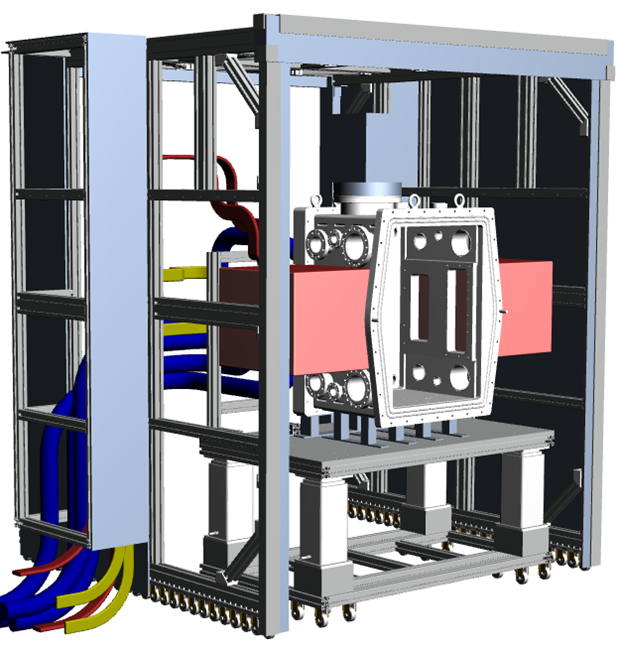}
\end{subfigure}
\caption{\label{fig:ba}  Left: shielding cabinet of the Big Amber. Right: a 3D CAD model rendering of the third cabinet containing the AGIPD vessel. Copyright The Author(s), published under the CC-BY 4.0 license (\href{http://creativecommons.org/licenses/by/4.0/}{http://creativecommons.org/licenses/by/4.0/})}
\end{figure}

The ambient test setup Big Amber is shown in figure \ref{fig:ba}, its shielding cabinet is $1.5$ m $\times$ 2 m $\times$
3.5 m large, such that it can house the 1~mega pixel detectors that are
presently foreseen to be used at the XFEL.EU.  The shielding cabinet is divided into three sections. 
The first two consist of movable breadboard tables on which the frame and the shielding of $3$ mm thick
lead is mounted. Sliding doors give access to the inside. Both sections 
can be moved away from the third, the shielding of which
reaches to the ground. One double door on the backside of the third section opens into a
clean room and allows detectors to be easily moved into the
setup once they are mounted on the detector support.
Figure~\ref{fig:ba} also shows a CAD rendering of the third cabinet containing
the vacuum vessel of the AGIPD  1~mega pixel detector mounted on the
detector support.

Of all setups Big Amber can provide the highest intensity 
due to the Seifert high power X-ray generator used. Tests requiring pulsed X-rays can be achieved using a high speed chopper, which produces pulses as short as $168$ ns .

\begin{figure}[b]
\includegraphics[width=0.44\textwidth,]{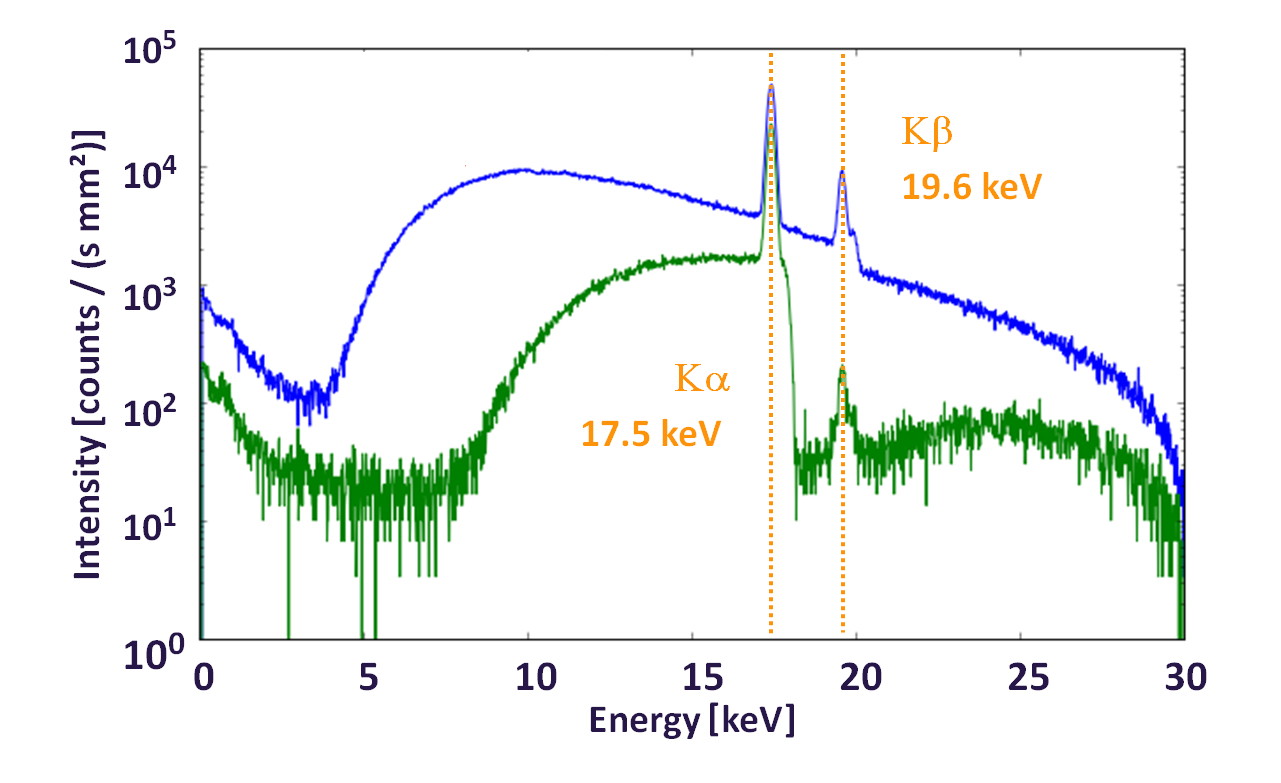}
\includegraphics[width=0.56\textwidth,]{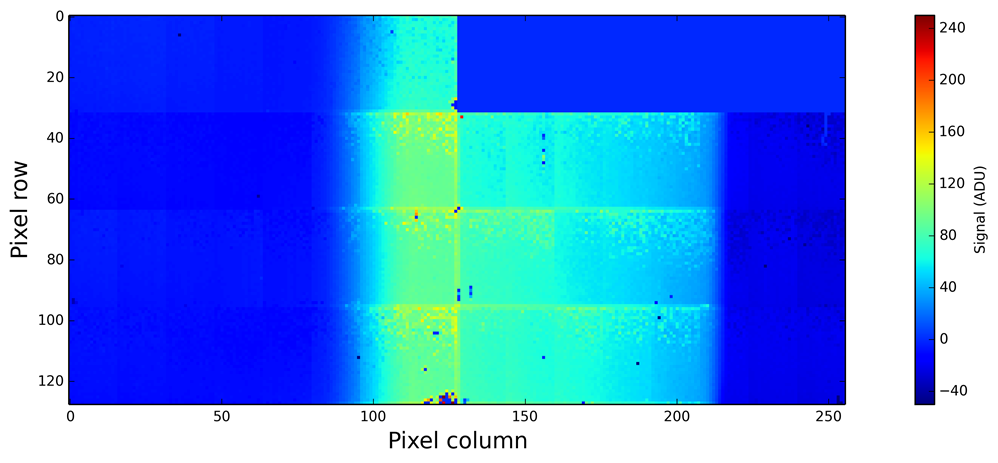}
\caption{\label{fig:mo} Left: X-ray emission spectrum of the Molybdenum tube without (blue) and with (green) a Zirconium filter.  Right: Response of an LPD super module sensor  illuminated with the Molybdenum tube at 50 kV and 40 mA at a distance of 0.5 m. The module was equipped with seven of eight possible tiles. Dead pixels are visible as well as the edges of the tiles where the pixels are 20$\%$ larger and therefore detect a higher signal. Copyright The Author(s), published under the CC-BY 4.0 license (\href{http://creativecommons.org/licenses/by/4.0/}{http://creativecommons.org/licenses/by/4.0/})}
\end{figure}

The Seifert source is operated with Molybdenum or
Copper anodes, at a voltage of up to $60$ kV and power
up to $2$ kW. X-ray tubes with different anodes are quickly exchangeable and give the opportunity to
perform measurements at higher X-ray intensities with different photon
energies. Filters to reduce the K$_{\beta}$ radiation are included in
the tube housing. Figure \ref{fig:mo} shows the X-ray emission spectra
of a Molybdenum tube with characteristic K lines at $17.5$ keV and
$19.6$ keV,  as well as the effect of  Zirconium as K-edge filter on the shape of the emission spectrum. In this material the binding energy of the electron in the K shell of the atom lies with $\approx$ 18 keV between the two peaks. Due to the rapid drop of the transmission coefficient at this energy~\cite{booklet}, the intensity of the  K$_{\beta}$ radiation can be reduced effectively together with the underlying
Bremsstrahlung continuum.

Count rates of $~10^6$ counts mm$^{-2}$ s$^{-1}$ can be provided with the
unfocused beam at a distance of $0.5$ m. With an opening angle of approximately 8 degrees the beam spot at this distance allows 
illumination of several thousand pixels simultaneously, as shown for a small prototype of the LPD in figure  \ref{fig:mo}, where offset corrected data from the sensor, irradiated with th Molybdenum X-ray tube, can be seen. 

A focusing polycapillary optic from IFG can be installed to increase the photon intensity delivered to a pixel. This provides a focal spot of less than $50 \mu$ m resulting in an intensity gain of up to $10^4$, depending on the photon energy, when compared to an unfocused beam. The commissioning of the optics is currently ongoing.

The Big Amber setup allows testing 1 Mpx Detectors with large-area illumination for flat field correction as well as scans of single pixels at medium or high photon energies.

\section{Vacuum setup with pulsed electron source PulXar}
The PulXar setup is currently being developed and will contain a pulsed electron source that is
capable of producing pulses of $50$-$150$ ns duration, within a $0.6$
ms burst followed by a $99.4$ ms gap \cite{infra}. A multi-target
anode will allow generation of characteristic X-rays at energies
between $0.3$ keV and $17.5$ keV. In addition a filter wheel and
polycapillary optics are foreseen.  

\begin{figure}[htbp]
\hspace{12mm}
\includegraphics[width=0.21\textwidth]{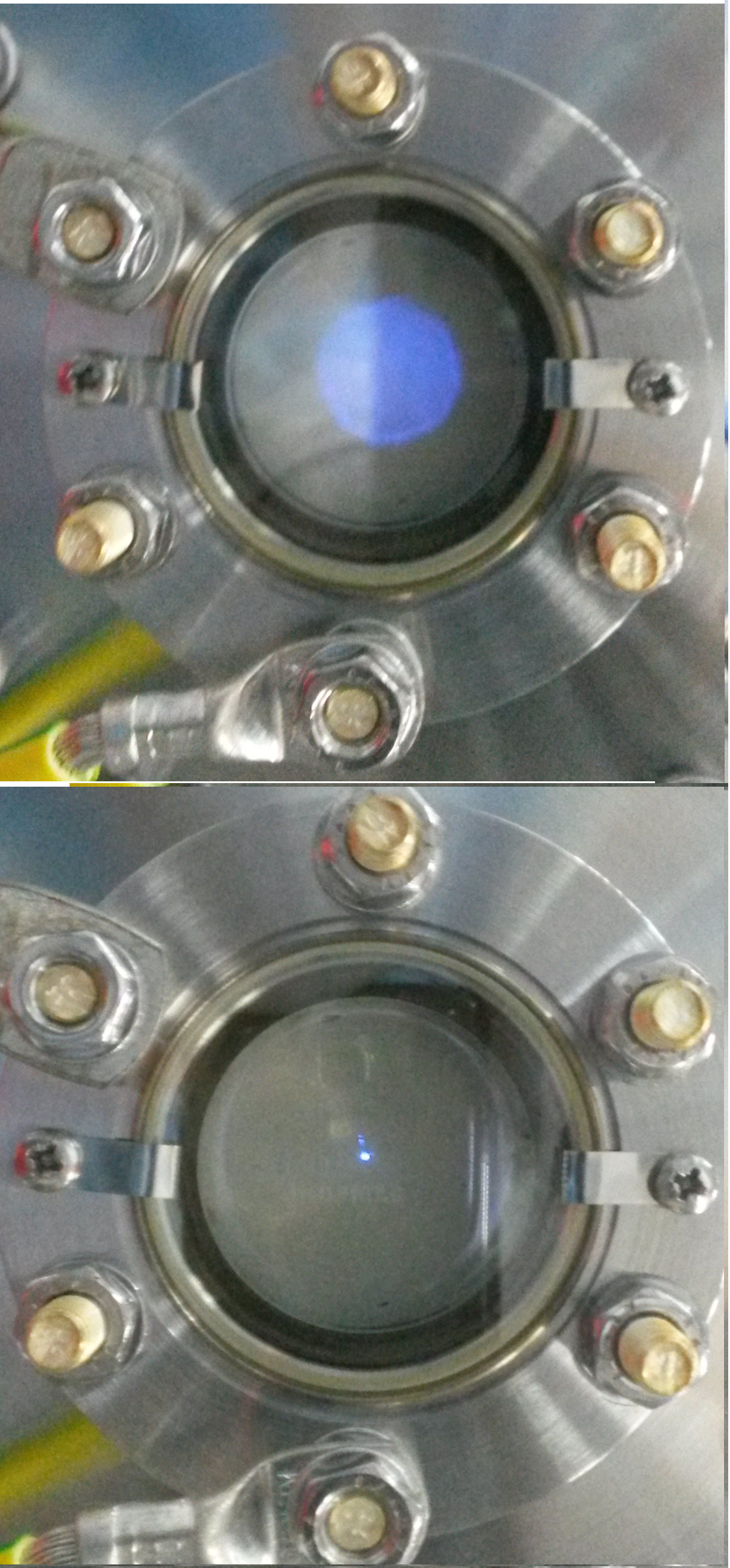}
\hspace{5mm}
\includegraphics[width=0.56\textwidth,]{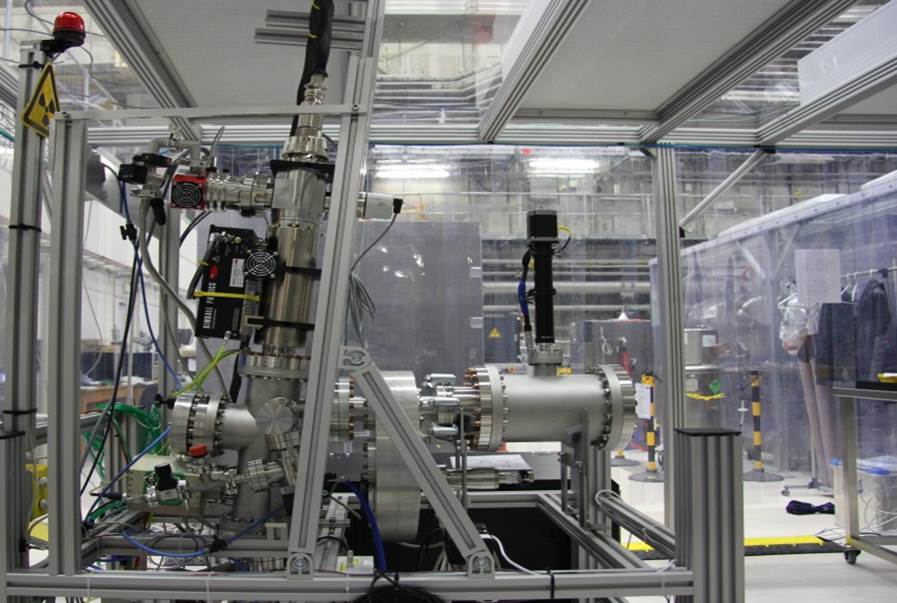}
\caption{\label{fig:pulxarv01} Left: beam spot of the unfocused beam (upper) and spot of focused beam (lower) with a lens current of $300$ mV. Right: Vessel of the pulsed X-ray source PulXar with filter wheel provided by the MPE combined with the pulsed electron gun from Kimball. Copyright The Author(s), published under the CC-BY 4.0 license (\href{http://creativecommons.org/licenses/by/4.0/}{http://creativecommons.org/licenses/by/4.0/})}
\end{figure}

The electron gun was
obtained from Kimball with a customized blanker section to provide the
necessary timing structure. With magnetic lenses the electron beam can
be focused on the target material to be used.  In
figure~\ref{fig:pulxarv01} the electron beam can be seen on a phosphor
screen attached to the beam outlet of the electron gun in case of an
unfocused and a focused beam.

While the design of the setup components is ongoing, as a testbed a
copy of the continuous multi target X-ray source, on loan from the
Max-Planck Institute for extraterrestrial Physics (MPE), was
assembled, commissioned and characterized.  As an intermediate step
towards the final PulXar setup the pulsed electron gun was combined
with a target and filter wheel assembly provided by MPE, as seen in
figure~\ref{fig:pulxarv01}. Commissioning of this setup is currently
in progress.
PulXar will provide the unique opportunity for timing tests with a pulse structure similar to the European XFEL. Due to the use of a target wheel with multiple different target materials it will be possible to provide a pulsed beam for any photon energy that is needed for tests with the XFEL detectors. 

\section{Summary}
While dedicated beam time at XFEL.EU for characterization, commissioning and
calibration of detectors will still be mandatory, e.g. to test the
full dynamic range of the $4.5$ MHz 2D cameras, the laboratory
setups described will provide the possibility to carry out a variety of
tests independent of beam time.
The two portable setups PHEOBE and Little Amber are available for basic tests with detectors of smaller sensor
size like the FastCCD and pnCCD. 
Big Amber has been used for tests with a smaller versions of AGIPD and LPD
and is ready for measurements with the 1~mega pixel versions of
these two detectors.
Synchronization and timing tests will be realized with
the customized pulsed electron source of the PulXar setup, that can
provide a similar timing structure as the XFEL burst mode.  While the
final version of the PulXar setup is still in the design phase, a
preliminary setup making use of the pulsed electron gun is under
commissioning.

\section{Acknowledgments}

The authors would like to thank the Max-Planck Institute for
extraterrestrial Physics for providing an X-ray source with filter
wheel for testing in our laboratory.


\end{document}